\documentclass[conference]{IEEEtran}
\IEEEoverridecommandlockouts

\usepackage{cite}
\usepackage{amsmath,amssymb,amsfonts}
\usepackage{algorithmic}
\usepackage{graphicx}
\usepackage{textcomp}
\usepackage{xcolor}
\def\BibTeX{{\rm B\kern-.05em{\sc i\kern-.025em b}\kern-.08em
    T\kern-.1667em\lower.7ex\hbox{E}\kern-.125emX}}

\usepackage{siunitx}
\newcommand\lowr[1]{_\textrm{#1}}

\newcommand\one[1]{\textcolor{black}{#1}}
\newcommand\two[1]{\textcolor{black}{#1}}
\newcommand\three[1]{\textcolor{black}{#1}}

\begin{document}

\title{Reconstructing Unobservable Temperature Fields via Simulation-Aided Intelligent Sensing\\
\thanks{This work has been supported by Silicon Austria Labs (SAL), owned by the Republic of Austria, the Styrian Business Promotion Agency (SFG), the federal state of Carinthia, the Upper Austrian Research (UAR), and the Austrian Association for the Electric and Electronics Industry (FEEI). }
}

\author{\uppercase{Monika Stipsitz}\authorrefmark{1},
\uppercase{Hèlios Sanchis-Alepuz}\authorrefmark{1}, 
\uppercase{Jacob Reynvaan}\authorrefmark{1}, 
and \uppercase{Silvester Sabathiel}\authorrefmark{1}}

\author{\IEEEauthorblockN{1\textsuperscript{st} Monika Stipsitz}
\IEEEauthorblockA{\textit{Embedded Systems} \\
\textit{Silicon Austria Labs GmbH}\\
Graz, Austria \\
monika.stipsitz@silicon-austria.com}
\and
\IEEEauthorblockN{2\textsuperscript{nd} Hèlios Sanchis-Alepuz}
\IEEEauthorblockA{\textit{Embedded Systems} \\
\textit{Silicon Austria Labs GmbH}\\
Graz, Austria \\
helios.sanchis-alepuz@silicon-austria.com}
\and
\IEEEauthorblockN{3\textsuperscript{rd} Jacob Reynvaan}
\IEEEauthorblockA{\textit{Power Electronics} \\
\textit{Silicon Austria Labs GmbH}\\
Graz, Austria \\
jacob.reynvaan@silicon-austria.com}
\and
\IEEEauthorblockN{4\textsuperscript{th} Silvester Sabathiel}
\IEEEauthorblockA{\textit{Embedded Systems} \\
\textit{Silicon Austria Labs GmbH}\\
Graz, Austria \\
silvester.sabathiel@silicon-austria.com}
}

\maketitle

\begin{table}[b] 
\footnotesize
\copyright~2026 IEEE. Personal use of this material is permitted. Permission from IEEE must be obtained for all other uses, in any current or future media, including reprinting/republishing this material for advertising or promotional purposes, creating new collective works, for resale or redistribution to servers or lists, or reuse of any copyrighted component of this work in other works. Presented at: IEEE International Instrumentation and Measurement Technology Conference (I2MTC), Nancy, France, 2026.
\end{table}

\begin{abstract}
Real-time monitoring of the temperature distribution within components and sub-structures is a challenging topic in many systems due to restrictions on feasible sensor locations. While machine learning (ML) proves a versatile tool in many applications, its adoption for high-resolution thermal monitoring is hindered by the availability of high-quality datasets for training. In this work, we propose a novel approach for generating datasets for industrial applications based on randomized physics-based simulations. We demonstrate the approach in a proof-of-concept hardware setup: A neural network (NN) trained only on such a synthetic dataset, is used to reconstruct the internal temperature field from sparse sensors embedded in the hardware. The NN-based reconstructions do not only outperform Kriging in robustness but also enable real-time inference, making the method suitable for online monitoring of otherwise unobservable thermal states.
\end{abstract}


\section{Introduction}

Monitoring thermal distributions is critical for system health in many applications such as electronic devices, fuel cells or battery packs, but full-field measurements are challenging due to inaccessible locations and practical limits on the number of embedded sensors.

Physics-based simulations \cite{Mo2025} and reduced-order models \cite{Aguado2015} can interpolate sparse measurements but are computationally intensive and require detailed knowledge of materials and boundary conditions (BC), limiting real-time applicability. Other methods rely on a combination of classical regression and statistics \cite{Ahn2022,shao2019bayesian}.
Recent ML approaches for thermal monitoring \cite{Hughes2023,Yule2025} mostly predict temperatures at a few locations where ground-truth data are available, highlighting that a key obstacle for industrial deployment is the lack of suitable high-quality data. Physics-informed neural networks (PINNs) \cite{Karniadakis2021} have been applied for full-field temperature reconstruction when no measurement data are available \cite{zhao2023physics,Manavi2023}, but they require precise knowledge of BC, material distribution, and heat sources, which is often unavailable in practical systems due to non-uniform reactions, component aging, or process variability.

In this work, we propose a strategy for generating high-quality training datasets for thermal monitoring based on randomized physics-based simulations. Using a proof-of-concept hardware setup, we show that the resulting randomized simulation dataset (RSD) enables the neural network to effectively combine prior knowledge learned from simulation with sparse sensor measurements to accurately reconstruct the temperature field.

The proposed RSD offers several advantages over classical measurement-based data collection: (1) it eliminates the need for costly and time-consuming dedicated experiments, which often yield biased datasets with limited coverage of operating conditions; (2) simulations provide arbitrary spatial and temporal resolution, including at inaccessible locations where sensors cannot be deployed, as demonstrated in this work by reconstructing an internal slice of a steel plate; (3) randomized simulation parameters reduce the requirement for exact system replication, allowing the model to remain robust under uncertain BC and manufacturing tolerances; and (4) unlike sparse and noisy measurements, synthetic simulation data inherently satisfy the governing physical laws, exposing the network to a broader physically consistent range of scenarios. Combined with a physics-informed objective, this improves generalization and robustness beyond purely data-driven learning.

\section{Materials \& Methods}

To demonstrate the proposed framework, we consider the reconstruction of the internal temperature field in a steel plate with complex surface channels. After introducing the hardware setup (Subsec.~\ref{sec:measurement_system}), we develop a simplified simulation model with randomized parameters (Subsec.~\ref{sec:simulation_model}) and generate a training dataset through a sequence of simulations. Finally, a neural reconstruction model is trained to ``interpolate'' in the space of temperature distributions given only the training dataset, and select the most likely temperature distribution given a sparse set of sensor values (Subsec.~\ref{sec:reconstruction_model}).

\subsection{Proof-of-concept system}
\label{sec:measurement_system}

\begin{figure}[htbp]
    \centerline{\includegraphics[width=0.4\textwidth]{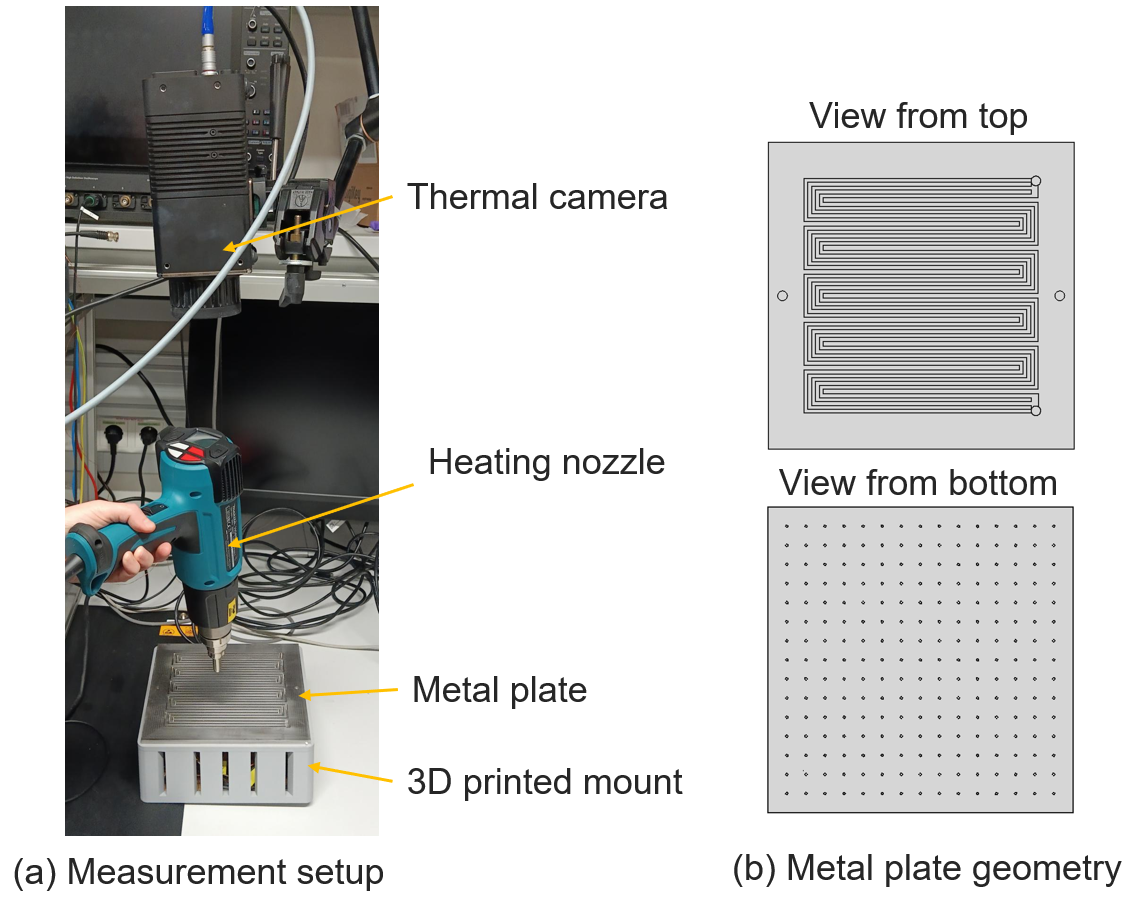}}
    \caption{(a) Hardware setup: For clarity the metal plate is shown without the anti-reflective coating applied for the application of the thermal camera. (b) The sensors are embedded in the metal plate at the holes shown in the view from bottom. To be able to study various sensor configurations, the plate contains more than the $N_s$ sensors that were actually placed.}
    \label{fig:measurement_setup}
\end{figure}

In our experimental setup, a steel plate (dimensions: $192 \times 192 \times \SI{10}{mm}$) is equipped with $N_s=17$ temperature sensors (TDK $10$K NTC thermistors B57550G1103+005) located at a depth of $z=\SI{2.6}{mm}$ within the plate (see Fig.~\ref{fig:measurement_setup}, left). The readout is performed using a voltage divider with a $10$K series resistor and via the ADC ports of two ESP32 boards. Measurements ares sampled at \SI{1}{Hz} and averaged over three samples. The sensors form an irregular grid with missing locations to emulate sensor failures. 
Heat is applied through a movable and tiltable hot-air nozzle ($\approx\SI{150}{\celsius}$) blowing at the top surface, producing complex and partially unknown heat-source patterns representative of realistic operating uncertainties.

In the following, we reconstruct the internal temperature field at $z=\SI{2.6}{mm}$ from the sparse sensor readings alone, without explicit knowledge of the heat input location, motion, or temporal history.

\subsection{Randomized Simplified Simulation Model}
\label{sec:simulation_model}

As noted above, the ML model is trained exclusively on simulation data, requiring a sufficiently large dataset to cover realistic operating conditions and temperature states. Generating hundreds of high-fidelity simulations would be computationally prohibitive; therefore, we employ a simplified surrogate model to reduce dataset generation time. To compensate for reduced physical detail, we introduce parameterized unknowns that are randomized across simulations.

This randomization serves two main purposes: (1) it captures uncertainties in system specifications, such as unknown BC, initial temperature distributions, and imperfect knowledge of the heat source characteristics; and (2) it avoids the need for exhaustive calibration to replicate individual experiments, since only plausible parameter ranges—derived from expert knowledge or a small number of test simulations—are required.

\textbf{Finite Element surrogate model:} For our proof-of-concept system, a high-fidelity model would require calibrated CFD simulations including the nozzle airflow. Instead, we employ a simplified heat-conduction FEM model, where the nozzle is approximated by Robin BC with heat transfer coefficient and external temperature varying according to a Gaussian distribution. The dataset is generated from multiple simulation sequences, each consisting of consecutive simulations restarted from the previous thermal state. Only the steel plate is explicitly modeled, while the epoxy mount and nozzle are represented as BCs; sensor holes are represented as epoxy (material properties in Tab.~\ref{tab:materials}).

\begin{table}[htbp]
\caption{Material properties used in FEM simulations}
\begin{center}
\begin{tabular}{|l|c|c|c|}
\hline
\textbf{Parameter} &
$\mathbf{k}$ &
$\mathbf{c\lowr{p}}$ &
$\mathbf{\rho}$ \\
\textbf{\textit{Unit}} & $\boldsymbol{\mathit{(W/(K m))}}$ & $\boldsymbol{\mathit{(J/(kg K))}}$ & $\boldsymbol{\mathit{(kg/m^{3})}}$ \\
\hline
Steel (316L) & 
15 & 450 & 8030 \\
Epoxy &
0.88 & 952 & 1682 \\
\hline
\end{tabular}
\label{tab:materials}
\end{center}
\end{table}

At the start of each simulation sequence, the whole plate is at a constant temperature $T\lowr{ini}$. 
On the top surface of the plate Robin BCs are assigned, given by
\begin{equation}
    f:= -k \frac{\partial T}{\partial n} = \alpha (T - T\lowr{ext}),
\end{equation}
where $k$ is the heat conductivity of the solid body, $\partial / (\partial n)$ denotes the differentiation in normal direction wrt the plate surface, $\alpha$ and $T\lowr{ext}$ are the heat transfer coefficient and external air temperature, respectively. 
To mimic the distributed heat injected into the plate via the air nozzle, the parameters $\alpha$ and $T\lowr{ext}$ of the Robin BCs are varied according to an elliptic gaussian distribution: Given a nozzle at position $(x_0, y_0)$ above the plate, the heat transfer coefficient is reduced from $\alpha\lowr{nozzle}$ at the center $r=\sqrt{(x-x_0)^2 + (y-y_0)^2}=0$ of the nozzle region, to $\alpha\lowr{air}$ at a cut-off radius $r\lowr{max}$. Similarly the external temperature is reduced from the nozzle temperature $T\lowr{nozzle}$ to the room temperature $T\lowr{air}$:
\begin{equation}
    \alpha(r) = \alpha\lowr{air} + (\alpha\lowr{nozzle} - \alpha\lowr{air}) \cdot \mathrm{e}^{-\left(\frac{x-x_0}{r_{0x}}\right)^2}\mathrm{e}^{-\left(\frac{y-y_0}{r_{0y}}\right)^2},
\end{equation}
and
\begin{equation}
    T\lowr{ext}(r) = T\lowr{air} + (T\lowr{nozzle} - T\lowr{air}) \cdot \mathrm{e}^{-\left(\frac{x-x_0}{r_{0x}}\right)^2}\mathrm{e}^{-\left(\frac{y-y_0}{r_{0y}}\right)^2}.
\end{equation}
Here, $(r_{0x}, r_{0y})$ are additional parameters that control how fast the flux parameters decrease in each direction with the distance $r$ from the center of the nozzle area. $r_{0y}$ is set via an ellipticity factor $\epsilon$: $r_{0y} = \epsilon \cdot r_{0x}$.
The applied flux BCs are thus given by
\begin{equation}
    f\lowr{nozzle}(r) = \alpha(r) \cdot (T - T\lowr{ext}(r))
\end{equation}
in the nozzle area ($r < r\lowr{max}$), and
\begin{equation}
    f\lowr{air} = \alpha\lowr{air} \cdot (T - T\lowr{air})
\end{equation}
in the remaining surface area. The air cooling at the side and bottom surfaces of the plate are modeled via Robin BCs with constant $\alpha\lowr{air}$ and $T\lowr{air}$.

\textbf{Randomization Approach:} To cover the range of temperature distributions that might occur in the monitored system (Subsec.~\ref{sec:measurement_system}), we generate simulations that individually do not replicate the system exactly but collectively span the space of possible temperature maps. Randomization parameters for the plate case are listed in Tab.~\ref{tab:parameter_ranges}, and for each individual simulation sampled from a uniform distribution with these ranges. The ranges were determined from domain knowledge and a simple calibration: the nozzle was placed above one embedded sensor for \SI{20}{s}. A few test simulations were used to determine parameter ranges, taking observed temperature variability and an uncertainty margin into account. Note, that in contrast to the naive approach of replicating the experiments with simulations, we require a much less stringent calibration process. Since the calibration only sets the range from which the actual parameters for each individual simulation are chosen, the agreement between simulation and measurement data does not have to be perfect.

\begin{table}[htbp]
\caption{Randomization parameter and ranges for the ML dataset}
\begin{center}
\begin{tabular}{|l|c|c|c|}
\hline
\textbf{Parameter} &
\textbf{Unit} &
\textbf{Minimum} &
\textbf{Maximum} \\
\hline
Planar nozzle position $(x_0, y_0)$ & mm & 0 & $192$ \\
\hline
$\alpha\lowr{air}$&
$W/(m^2 K)$&
15 & 25 \\
$T\lowr{air}$&
\si{\celsius} &
17 & 41 \\
$\alpha\lowr{nozzle}$&
$W/(m^2 K)$&
150 & 350 \\
$T\lowr{nozzle}$&
\si{\celsius} &
 100 & 160 \\
Gaussian width $r_{0x}$&
$mm$&
30 & 70 \\
Elliptic factor $\epsilon$ & $-$ & 0.5 & 1.0 \\
nozzle velocity $v\lowr{nozzle}$& $mm/s$ & 6 & 20 \\
(in-case of movement) & & & \\
\hline
Time per simulation $t_i$ & $s$ & 1 & 180 \\
Simulations per sequence $N$ & $-$ & 2 & 30 \\
\hline
\end{tabular}
\end{center}
\label{tab:parameter_ranges}
\end{table}

\begin{figure}[htbp]
    \centerline{\includegraphics[width=\linewidth]{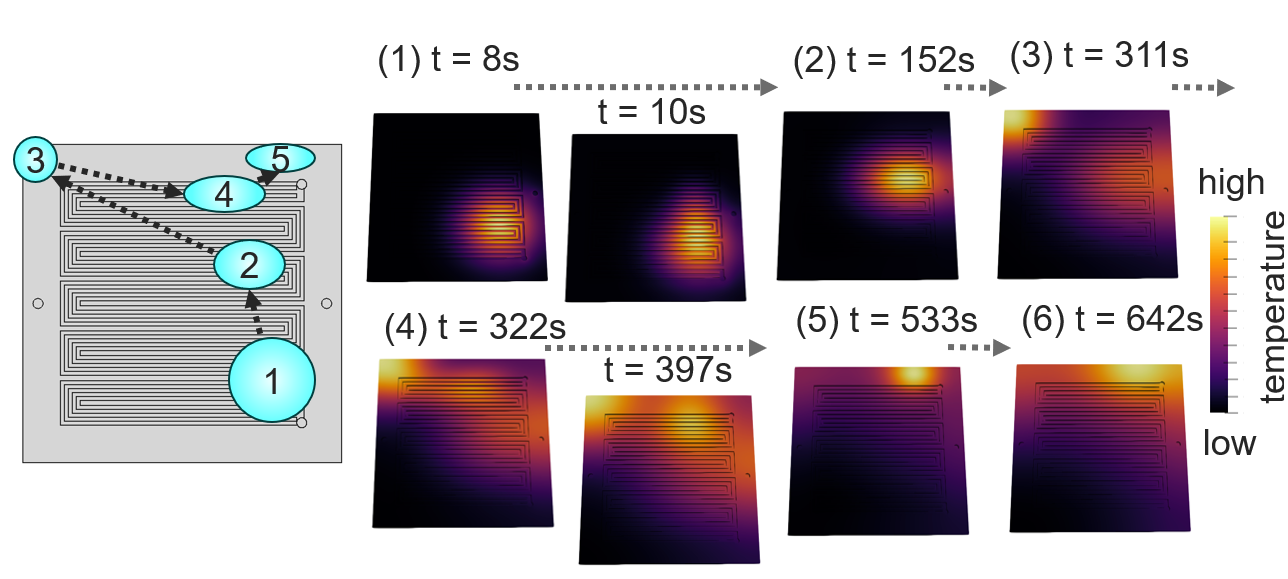}}
    \caption{Example for one simulation sequence consisting of multiple different heating simulations at locations (1-5) and final cooling period (point 6). The location and movement of the nozzle over the surface is shown in the left figure, the shape of ellipse is an approximation of the parameterized Gaussian at the given simulation instance.}
    \label{fig:dataset_example}
\end{figure}

\textbf{Dataset specifications:} In the previous paragraphs we detailed how the parameters for one distinct static heating process are determined. To match the concrete use case, we require that the ML dataset covers two additional scenarios: (1) nozzle movement across the plate, and (2) system cooling after the nozzle is turned off. To include these scenarios, each simulation consists of several segments: starting from a homogeneous initial temperature $T\lowr{ini}=T\lowr{air}$, a random nozzle position $(x_0,y_0)$ is selected and simulated for $t_0$s. The nozzle then moves to a new random position $(x_1,y_1)$ over $t_1$s with a random velocity $v\lowr{nozzle}$, followed by a segment at the constant position for $t_2$s. This process is repeated $N$ times, ending with a cooling period of $t_N$s. Simulations are sampled at \SI{1}{Hz}. An example sequence is shown in Fig.~\ref{fig:dataset_example}.

A total of 240 simulation sequences were generated. From each segment, two random time steps were selected and resampled to a $64 \times 64$ grid, yielding $10464$ 2D images. The dataset workflow was fully automated in Python using open-source tools: CAD design in FreeCAD, meshing with gmsh \cite{gmsh}, FEM simulations with ElmerFEM \cite{malinen2013elmer}, and postprocessing to extract sensor inputs and 2D temperature maps as labels.

\subsection{Reconstruction model}
\label{sec:reconstruction_model}

For monitoring the internal temperature distribution at a given time instance $t$, we reconstruct the full 2D heat map of $64 \times 64$ points based on the $N$ sensor values at time $t$. In principle, this could be achieved with a number of different methods, for simplicity, we chose a 2D NN in this work. However, for comparison we also show reconstructed temperature maps based on the classical method Kriging (\texttt{UniversalKrigging} from python module \texttt{pykrige} with standard parameters).

The specific NN architecture applied in this work is derived from \cite{sabathiel2024} 
: The 3 input channels are
\begin{enumerate}
    \item A masked 2D array of the temperature at the sensor positions at the current time step $T(t_i)$
    \item A masked 2D array of the temperature difference at the sensor positions from the current to the previous timestep $\Delta T = T(t_i) - T(t_{i-1})$
    \item A binary 2D mask indicating the locations of the sensors.
\end{enumerate}
The NN architecture consists first of a learned spatial propagator to fill the input temperature field with values interpolated from the given sensor values (with $10$ linear reconstruction steps). These layers are followed by a U-Net structure to reconstruct the full 2D temperature field at the internal slice (see \cite{sabathiel2024} for details). At last, the actual sensor values are inserted into the reconstruction map.

\three{The focus of this work is the generation of physics-conforming training data rather than ML architecture innovation. The proposed framework is model-agnostic, and the FEM-based data generation and uncertainty-aware validation procedures can be applied to alternative architectures such as graph neural networks or transformer-based models, as demonstrated in prior work \cite{LI2025109393, HU2025125033}. A detailed architectural comparison is therefore left for future work.}

\textbf{Increase robustness:} To make the model more robust, multiplicative Gaussian noise ($\mu=1$, $\sigma=0.005$) is added to the first two input channels during training to emulate noise and limited accuracy of the hardware sensors. The noise is sampled pixel-wise at each forward pass. Additionally, the model is trained with $6$–$8$ randomly removed sensors to enhance resilience against sensor failures.

\textbf{Training loss:} Following \cite{sabathiel2024}, three distinct loss terms were applied: (1) The standard MSE loss of the temperature pixels $L_T$, (2) the MSE loss of the temperature gradient $L_G$ computed via discrete finite differences from the 2D temperature fields, and (3) a physics-based loss term $L\lowr{H}$. In contrast to \cite{sabathiel2024}, $L\lowr{H}$ was based on the transient heat equation rather than the steady-state one:
\begin{equation}
    L\lowr{H}(T'_t, T_t, T_{t-1}) = \frac{1}{N} \left|\left| \sum_{\text{pixels}} \left(C_H(T_t, T_{t-1}) - C_H(T'_t, T_{t-1})\right)\right|\right|, \label{eq:loss}
\end{equation}
with $N$ the total number of pixels in the image. $T$ and $T'$ indicate that the temperate distribution obtained from the FEM simulation and from the reconstruction NN are used for the evaluation, respectively. The deviation from the heat equation constraint is given by
\begin{equation}
        C_H(T_{t}, T_{t-1}) = \frac{\partial T}{\partial t} - \frac{1}{\rho c\lowr{p}} \nabla \cdot (k \nabla T), \label{eq:heat_eq}
\end{equation}
where $k$, $c\lowr{p}$, and $\rho$ denote the spatially varying heat conductivity, specific heat capacity and density, respectively. Note, that in Eq.~\ref{eq:loss}, the NN based estimation is only used for the current time step $t$, while the FEM simulation data is taken for the temperature fields at the previous time step. The spatial grid used for the evaluation of the NN is too coarse to allow accurate discrete differentiation in case of varying $k$. Thus, $L\lowr{H}$ is only applied to grid points with constant $k$. Then, the differential operations can be evaluated using discrete finite differences (time increment $\Delta t = \SI{1}{s}$, spatial discretizations $\Delta x = \Delta y = \SI{3}{mm}$ (in-plane), $\Delta z = \SI{0.5}{mm}$ (out-of-plane)).

In contrast to PINNs, the heat equation loss $L\lowr{H}$ as derived from the discrete temperature maps is not accurate enough to be used as driving force during the NN training. Instead it is used as a regularizing term which was found in previous works to help in generalization and to speed-up convergence during the training process \cite{stipsitz2022approximating,sabathiel2024}. Thus, the full loss is given by $L =  L_T + \alpha_G \cdot L_G + \alpha_H \cdot L\lowr{H}$,
with the scaling factors $\alpha_G = 0.1$ and $\alpha_H = 10^{-5}$. 
Training was performed exclusively on the simulation dataset. \three{The FEM simulations were split into disjoint training (80\%) and test (20\%) datasets, with the test set used exclusively for quantitative evaluation.} The training was continued for $7000$ epochs using a step scheduler for the learning rate with batch size of $20$ (initial learning rate $0.001$, step of $100$, multiplicative decay factor $0.9$).

\section{Results}

\begin{figure}[htbp]
    \centerline{\includegraphics[width=0.5\textwidth]{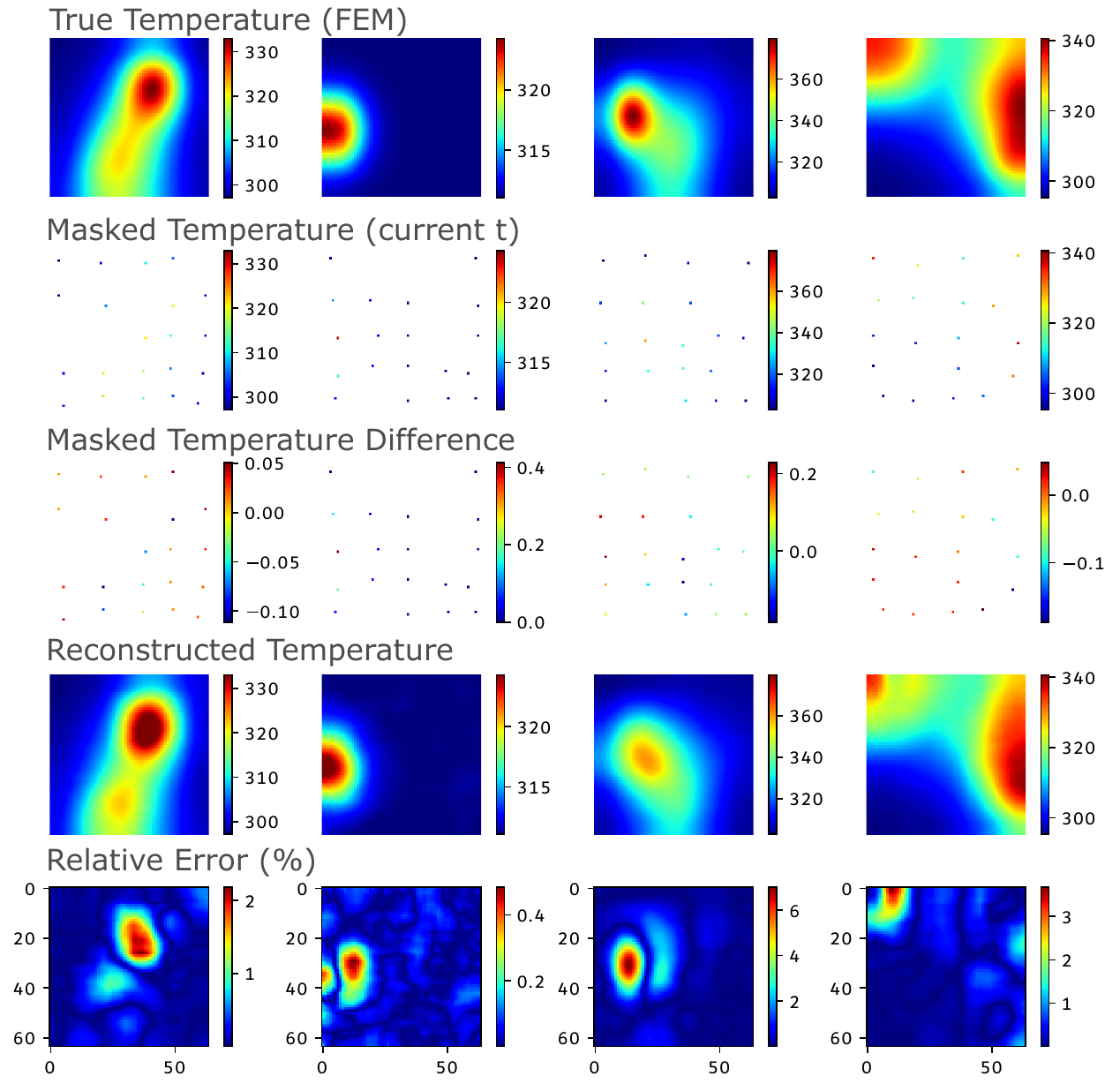}}
    \caption{Internal temperature reconstruction for selected samples \three{from unseen FEM test cases}. The NN reconstructs temperature maps (row 4) from masked input images $T_t$ and $\Delta T$ (rows 2–3) that closely match the ground truth (row 1).}
    \label{fig:simulation_data}
\end{figure}

Fig.~\ref{fig:simulation_data} illustrates the variability of temperature maps that the NN learns to reconstruct: The NN accurately reconstructs temperature fields at sensor height and generalizes to variable sensor numbers and configurations within the training space (see variable input masks in row 2).

\begin{figure}[htbp]
        \centering
        \begin{minipage}{0.12\textwidth}
            \centerline{Sensors $T_t$}
        \end{minipage}%
        \begin{minipage}{0.12\textwidth}
            \centerline{NN}
        \end{minipage}%
        \begin{minipage}{0.12\textwidth}
            \centerline{Kriging}
        \end{minipage}%
        \begin{minipage}{0.12\textwidth}
            \centerline{Surface}
        \end{minipage}\\

        \vspace{2pt}
        \rule{0.45\textwidth}{0.4pt}
        \vspace{4pt}
        
        \rotatebox{90}{\hspace{0.5cm}$t = \SI{1}{s}$}%
        \hspace{0.1cm}
        \includegraphics[width=0.45\textwidth]{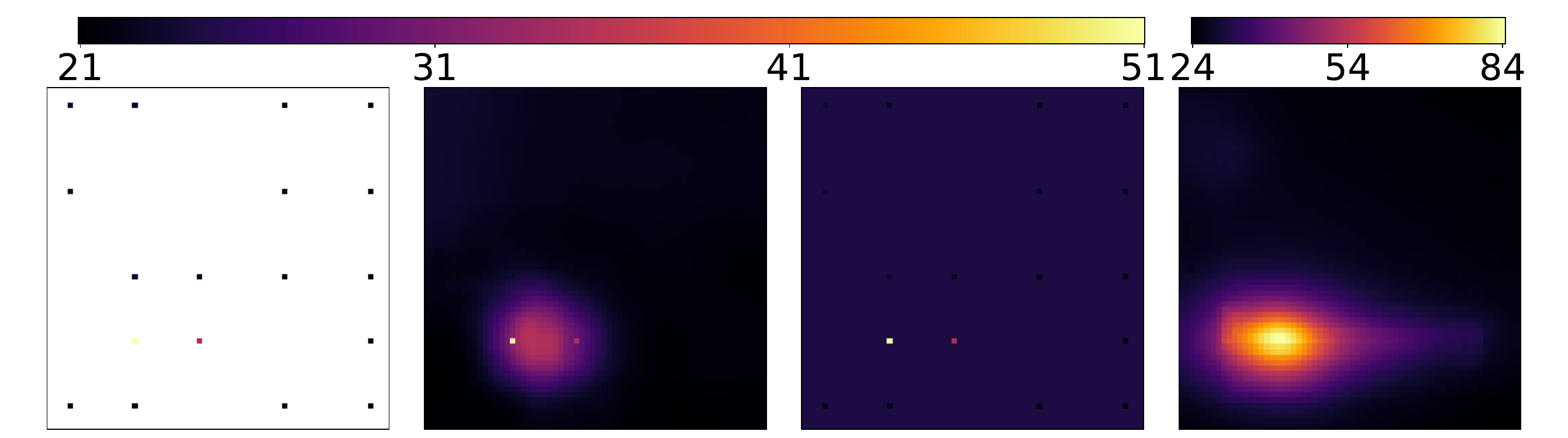}\\
         \rotatebox{90}{\hspace{0.25cm}$t = \SI{100}{s}$}%
         \hspace{0.1cm}
        \includegraphics[width=0.45\textwidth]{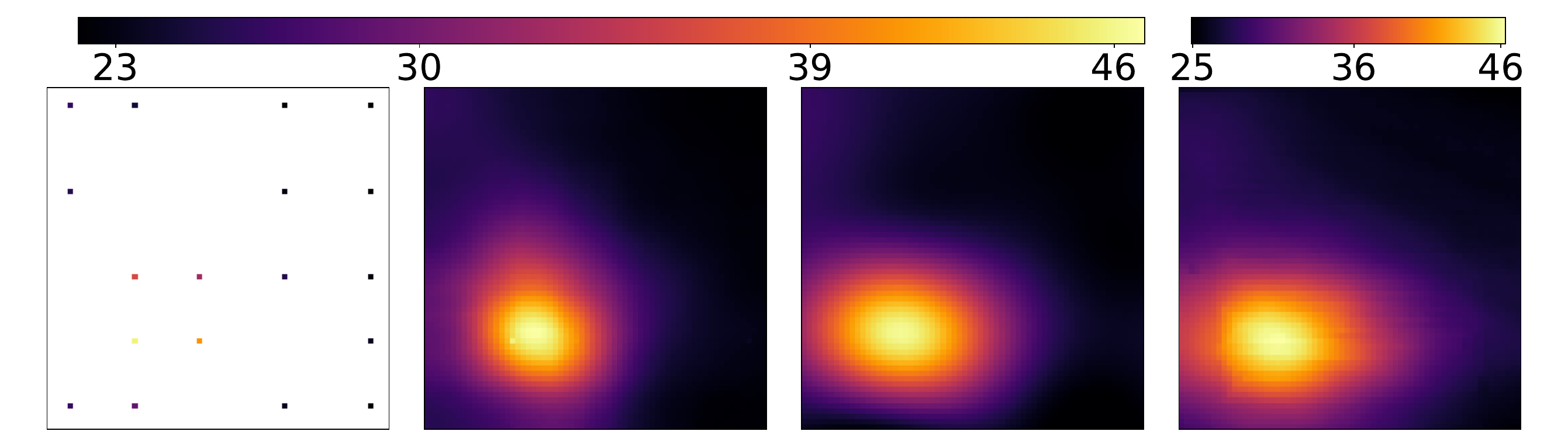}\\

        \vspace{2pt}
        \rule{0.45\textwidth}{0.4pt}
        \vspace{4pt}

        \rotatebox{90}{\hspace{0.5cm}$t = \SI{1}{s}$}%
        \hspace{0.1cm}
        \includegraphics[width=0.45\textwidth]{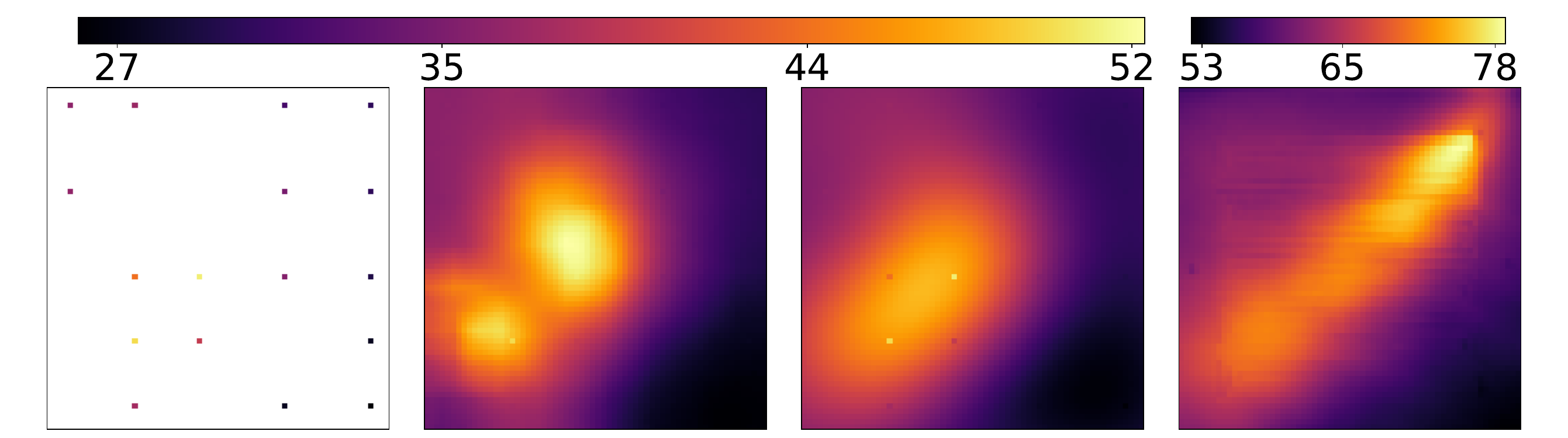}\\
        \rotatebox{90}{\hspace{0.5cm}$t = \SI{60}{s}$}%
        \hspace{0.1cm}
        \includegraphics[width=0.45\textwidth]{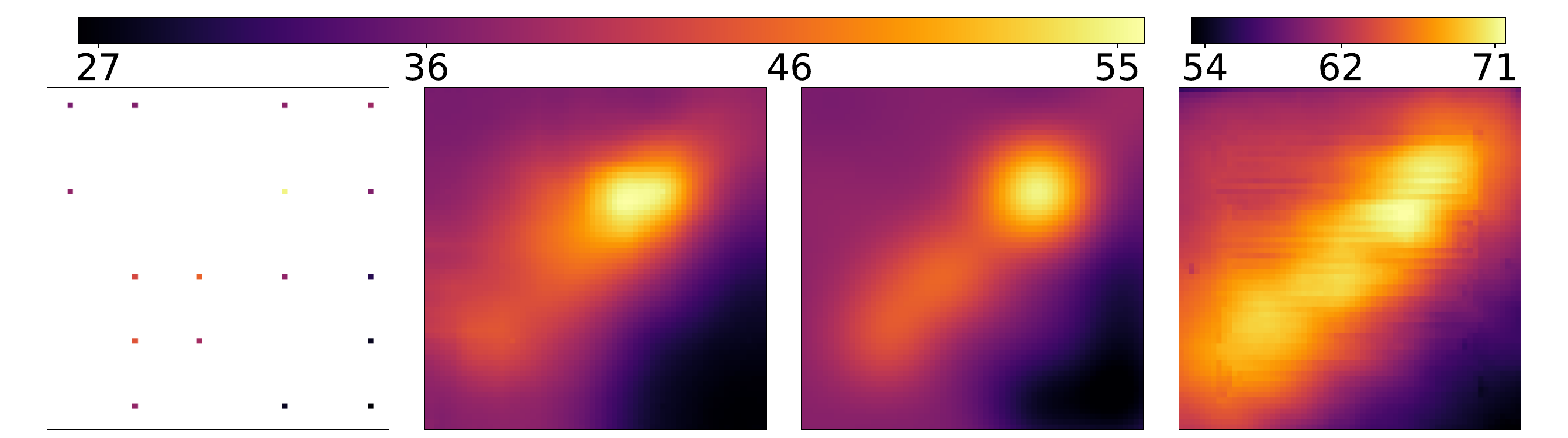}\\  

        \vspace{2pt}
        \rule{0.45\textwidth}{0.4pt}
        \vspace{4pt}

        \rotatebox{90}{\hspace{0.5cm}$t = \SI{1}{s}$}%
        \hspace{0.1cm}
        \includegraphics[width=0.45\textwidth]{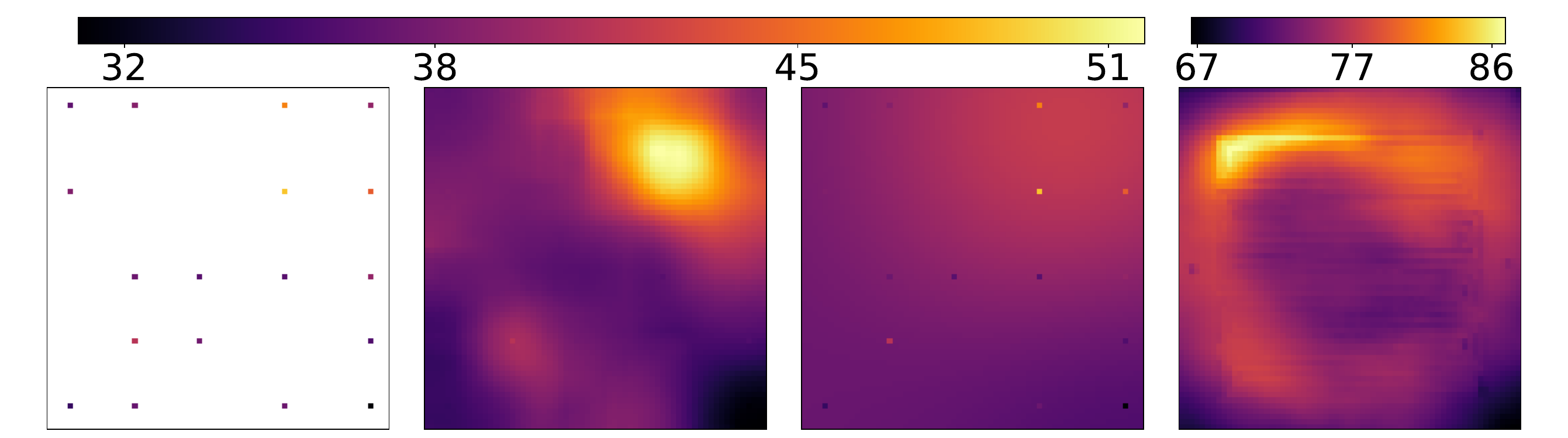}\\
        \rotatebox{90}{\hspace{0.5cm}$t = \SI{100}{s}$}%
        \hspace{0.1cm}
        \includegraphics[width=0.45\textwidth]{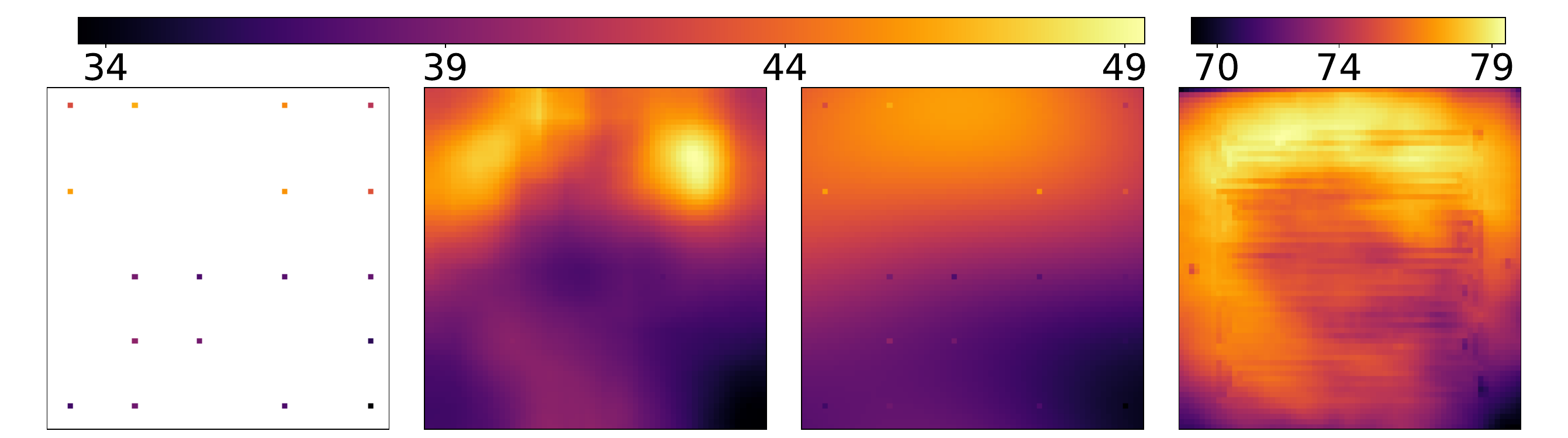}

    \caption{Static heating from uniform initial temperature state (rows 1 \& 2), diagonal heating path (rows 3 \& 4) and circular heating path (rows 5 \& 6) from a non-uniform initial temperature state. Each row shows the temperature (in \textdegree C, colorbar above the image) masked sensor data (col.~1), NN reconstruction (col.~2), Kriging reconstruction (col.~3), and thermal camera image (col.~4).}
    \label{fig:combined_res}
\end{figure}

After training, the NN was applied to reconstruct the temperature distribution in the hardware demonstrator (Sec.~\ref{sec:measurement_system}) from the recorded sensor data at sensor height. Direct validation of internal temperatures was not possible, so thermal camera images (InfraTec VarioCam® HD and IR camera) were collected above the plate to assess plausibility. Three measurement experiments were conducted (Fig.~\ref{fig:combined_res}): (1) static heating for $t \approx \SI{30}{s}$ from a uniform initial temperature, (2) diagonal nozzle movement from lower-left to upper-right starting from a non-uniform state, and (3) counter-clockwise circular nozzle movement starting near the lower-left corner. For experiments (2) and (3), non-uniform initial temperatures were generated by random heating and cooling sequences over ~$\SI{10}{min}$. 

\begin{table*}[htbp]
\caption{\three{Ablation study evaluated on held-out FEM test systems}}
\begin{center}
\begin{tabular}{|l|c|c|c|c|c|}
\hline
\textbf{Ablation} & $\boldsymbol{SCE_{\mathrm{RMSE}}}$ & \textbf{Mean}$\boldsymbol{(\sqrt{L_T})}$ &  \textbf{Max}$\boldsymbol{(\sqrt{L_T})}$ & \textbf{Mean}$\boldsymbol{(L\lowr{H})}$ & \textbf{Max}$\boldsymbol{(L\lowr{H})}$  \\
   & \textbf{\textit{(K)}} & \textbf{\textit(K)} & \textbf{\textit(K)} & \textbf{\textit(K/s)} & \textbf{\textit(K/s)}  \\
  \hline
Original & 2.02 & 0.81 & 5.89 & 0.02 & 0.09 \\ 
\hline
$\alpha_H = 0$ & 2.01 & 0.80 & 5.98 & 0.06 & 0.19 \\ 
\hline
1 failed sensor & 2.47 & 0.97 & 16.73 & 0.02 & 0.09 \\ 
2 failed sensors & 2.99 & 1.22 & 11.11 & 0.02 & 0.15 \\
3 failed sensors & 3.66 & 1.51 & 18.17 & 0.02 & 0.14 \\
4 failed sensors & 4.49 & 1.87 & 16.36 & 0.03 & 0.18 \\
\hline
Noise ($\sigma=\SI{0.5}{K}$) & 2.43 & 1.1 & 11.46 & 0.02 & 0.13 \\
Noise ($\sigma=\SI{1.14}{K}$) & 3.65 & 1.83 & 12.12 & 0.02 & 0.1  \\
Noise ($\sigma=\SI{2.0}{K}$) & 5.4 & 2.92 & 17.55 & 0.04 & 0.18  \\
\hline
\end{tabular}
\label{tab:ablation}
\end{center}
\end{table*}

\textbf{Ablation study:} \three{To assess the robustness of the proposed simulation-aided sensing framework, we conducted an ablation study on a held-out FEM test dataset (Tab.~\ref{tab:ablation}). Four configurations were evaluated: (i) the full model with all loss terms, (ii) a model trained without the heat-equation loss, (iii) the full model evaluated with 1–4 randomly failed sensors, and (iv) the full model evaluated under additive sensor noise. Performance was quantified using a sensor-consistency RMSE computed as the average over a leave-one-sensor-out protocol ($\textrm{SCE}\lowr{RMSE} = \sqrt{1/N_s\sum_{i=1}^{N_s} (T'_{-i} - T_i)^2}$, with the subscript $-i$ indicating that sensor $i$ is left-out), spatial reconstruction accuracy measured by the mean and maximum RMSE $\sqrt{L_T}$ over the temperature field, and physical consistency quantified by the mean and maximum physics residual $L\lowr{H}$ of the governing heat equation. This ablation isolates the impact of physics-based regularization and measurement perturbations while keeping the network architecture fixed, thereby directly evaluating the proposed data generation and training strategy.}

\begin{figure}[htbp]
        \centering
        \begin{minipage}[c]{0.25\textwidth}
            \centerline{Mean $T\lowr{NN}$}
        \end{minipage}%
        \begin{minipage}[c]{0.25\textwidth}
            \centerline{$\sigma\lowr{NN}$}
        \end{minipage} \\
        \begin{minipage}[c]{0.25\textwidth}
            \includegraphics[width=\textwidth]{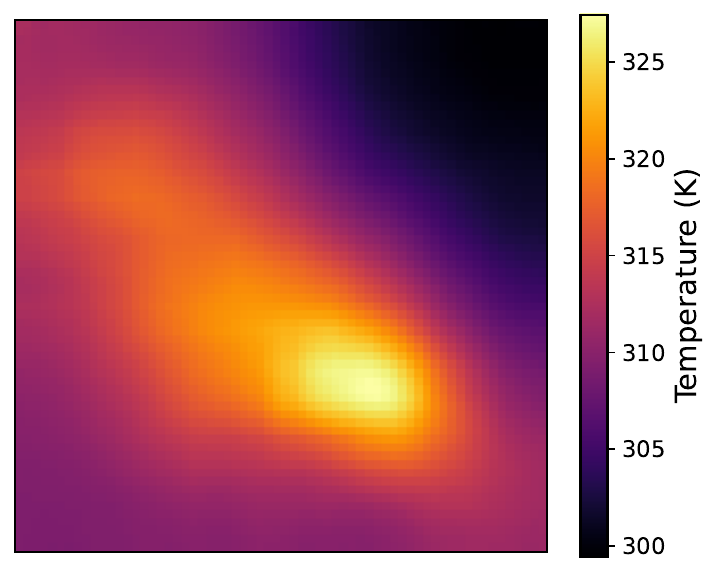}
        \end{minipage}%
        \begin{minipage}[c]{0.25\textwidth}
            \includegraphics[width=\textwidth]{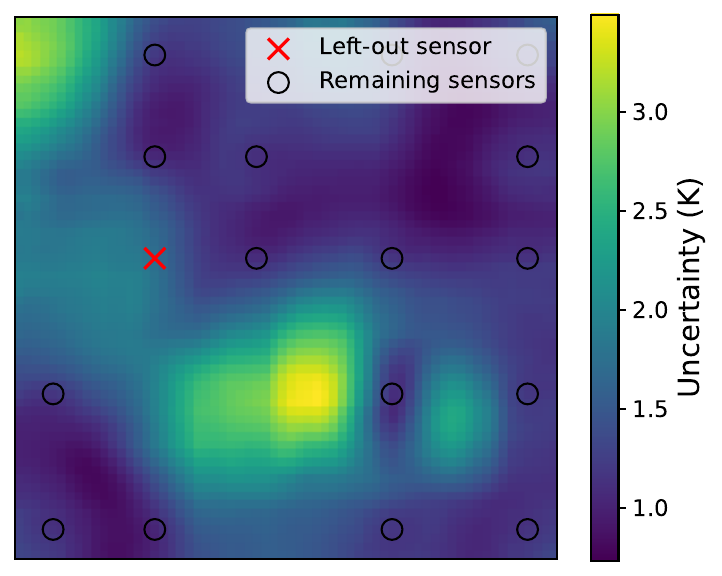}
        \end{minipage}\\
        \caption{\one{Mean reconstructed temperature map $T\lowr{NN}$ and corresponding uncertainty distribution $\sigma\lowr{NN}$ for the sample with the near-median $z_i$.}}
    \label{fig:uncertainty_map}
\end{figure}

\textbf{Measurement uncertainty \& compatibility:} \one{Related I\&M studies on sparse-sensor thermal field reconstruction remain limited; therefore, we adopt standard metrology frameworks (VIM/GUM) for uncertainty evaluation and compatibility assessment. The uncertainty of the temperature sensing system was evaluated by propagating the sensing chain uncertainties to the temperature domain. Each node uses a calibrated NTC thermistor in a voltage divider, read by a calibrated ADC configured as a current source. Thermistor nominal resistances were individually calibrated, while the $\beta$-parameter was taken from the datasheet ($\beta = \SI{3492}{K} \pm \SI{1}{\%}$ \cite{tdk_ntc_datasheet}). ADC noise ($\sigma_V = \SI{37}{mV}$ for ESP32-S3 \cite{esp32_adc_uncertainty}) and the $\beta$-parameter uncertainty were propagated via first-order sensitivity analysis. Assuming independent sources, the combined standard uncertainty is $\sigma\lowr{Sensor} \approx \SI{1.14}{K}$.
Following \cite{jcgmgum1,oiml_g101_supp1}, the NN is treated as a virtual measurement, and input uncertainties are propagated via Monte Carlo (MC) simulations. Gaussian noise ($\mu=0$, $\sigma=\sigma\lowr{Sensor}$) is added to sensor inputs, and $M=500$ MC runs estimate the NN output uncertainty $\sigma\lowr{NN}$. A leave-one-out strategy ensures independence: for each sensor, the temperature field is reconstructed using the remaining $N-1$ sensors, and the reconstructed value $T\lowr{NN}$ is compared with the measured $T\lowr{Sensor}$. The normalized compatibility index is defined as $z_i = |T\lowr{NN} - T\lowr{Sensor}| / \sqrt{\sigma\lowr{Sensor}^2 + \sigma\lowr{NN}^2}$. Fig.~\ref{fig:uncertainty_map} illustrates the mean reconstructed maps and uncertainty for the near-median $z_i$. Compatibility is assessed using the standard criterion for independent measurements: $|T\lowr{NN} - T\lowr{Sensor}| \le 2 \sqrt{\sigma\lowr{Sensor}^2 + \sigma\lowr{NN}^2}$ (with coverage factor $k=2$). Table~\ref{tab:compatibility} summarizes the results for the measurement scenarios in Fig.~\ref{fig:combined_res}.}

\begin{table}[h]
\caption{\one{Compatibility analysis}}
\begin{center}
\begin{tabular}{|l|c|c|c|}
\hline
\textbf{Sample} & \textbf{Compatibility rate (\%)} & \textbf{Mean} $\boldsymbol{z_i}$ & \textbf{Max} $\boldsymbol{z_i}$  \\ 
\hline
Static ($t=\SI{1}{s}$) & 88.9 & 1.1 & 10.3 \\ 
Static ($t=\SI{100}{s}$) & 94.4 & 0.7 & 2.5  \\ 
Diagonal ($t=\SI{1}{s}$) & 82.4  & 1.1 & 5.8  \\ 
Diagonal ($t=\SI{60}{s}$) & 70.6 & 1.2 & 6.0  \\ 
Circle ($t=\SI{1}{s}$) & 77.8 & 1.4 & 3.2  \\ 
Circle ($t=\SI{100}{s}$) & 94.4 & 0.9 & 2.4  \\ 
\hline
\end{tabular}
\label{tab:compatibility}
\end{center}
\end{table}

\section{Discussion}

The NN reconstructions produce qualitatively reasonable temperature maps, capturing non-uniform distributions and small heated regions better than Kriging, and are robust to individual sensor failures even for unseen configurations (Fig.~\ref{fig:combined_res}). Deviations from surface thermal camera images may reflect internal structural details or heat conduction\ delays. \three{The ablation study shows that including the physics-based loss improves physical plausibility; while noise and sensor failures increase RMSE, physics residuals remain comparatively bounded.} \one{The compatibility analysis indicates that predictions agree with sensor measurements, with mean normalized deviations around one standard deviation and compatibility rates of 71–94\%.}

\three{Although this study focuses on 2D temperature reconstruction from sparse sensors, previous work demonstrated the feasibility of simulation-aided reconstruction for predicting temperature evolution in complex 3D geometries with multiple materials and heat sources \cite{stipsitz2022approximating, helios2022graph}. These approaches did not consider sparse sensors or monitoring applications, but suggest that extending the current framework to higher-dimensional and heterogeneous systems is feasible.}

\section{Conclusion}

\two{In this work, we presented a simulation-aided framework for reconstructing internal (partially unobservable) temperature fields from sparse sensors. Trained exclusively on randomized physics-based simulations, the approach produces physically consistent reconstructions that are robust to sensor noise, failures, and sparsity, and better capture localized and non-uniform heating patterns than classical interpolation methods. Compatibility analysis confirms that predicted fields align with measured sensor values within expected uncertainty bounds. While demonstrated on a 2D steel plate, extending the framework to 3D geometries, heterogeneous materials, and alternative sensing strategies remains future work. Such extensions could enable simulation-aided training for more complex monitoring scenarios with sparse or partially unobservable sensors.}

\bibliographystyle{./IEEEtran}
\bibliography{literature}

@INPROCEEDINGS{Mo2025,
  author={Mo, Xianghao and Linares, Daniel Ríos and Ramos, Regina and Vasić, Miroslav},
  booktitle={2025 IEEE Applied Power Electronics Conference and Exposition (APEC)}, 
  title={FPGA-based Hybrid Simulator for Real-Time 3-D Temperature Monitoring of Power Converters}, 
  year={2025},
  pages={1375-1382},
  doi={10.1109/APEC48143.2025.10977068}}

@article{Aguado2015,
author = {Aguado, José V. and Huerta, Antonio and Chinesta, Francisco and Cueto, Elías},
title = {Real-time monitoring of thermal processes by reduced-order modeling},
journal = {International Journal for Numerical Methods in Engineering},
volume = {102},
number = {5},
pages = {991-1017},
doi = {10.1002/nme.4784},
year = {2015}
}

@article{Ahn2022,
  author    = {Ahn, C.-U. and Oh, S. and Kim, H.-S. and Park, D. I. and Kim, J.-G.},
  title     = {Virtual Thermal Sensor for Real-Time Monitoring of Electronic Packages in a Totally Enclosed System},
  journal   = {IEEE Access},
  volume    = {10},
  pages     = {50589--50600},
  year      = {2022},
  doi       = {10.1109/ACCESS.2022.3174208}
}

@article{shao2019bayesian,
  author    = {Shao, W. and Ge, Z. and Yao, L. and Song, Z.},
  title     = {Bayesian nonlinear Gaussian mixture regression and its application to virtual sensing for multimode industrial processes},
  journal   = {IEEE Transactions on Automation Science and Engineering},
  volume    = {17},
  number    = {2},
  pages     = {871--885},
  year      = {2019},
  publisher = {IEEE}
}

@article{Hughes2023,
title = {Real-time temperature prediction of electric machines using machine learning with physically informed features},
journal = {Energy and AI},
volume = {14},
pages = {100288},
year = {2023},
issn = {2666-5468},
doi = {10.1016/j.egyai.2023.100288},
author = {Ryan Hughes and Thomas Haidinger and Xiaoze Pei and Christopher Vagg}
}

@Article{Yule2025,
AUTHOR = {Yule, Lawrence and Harris, Nicholas and Hill, Martyn and Zaghari, Bahareh},
TITLE = {An Experimental Study of Machine-Learning-Driven Temperature Monitoring for Printed Circuit Boards (PCBs) Using Ultrasonic Guided Waves},
JOURNAL = {NDT},
VOLUME = {3},
YEAR = {2025},
NUMBER = {1},
ARTICLE-NUMBER = {1},
ISSN = {2813-477X},
DOI = {10.3390/ndt3010001}
}

@article{zhao2023physics,
  author    = {Zhao, X. and Gong, Z. and Zhang, Y. and Yao, W. and Chen, X.},
  title     = {Physics-informed convolutional neural networks for temperature field prediction of heat source layout without labeled data},
  journal   = {Engineering Applications of Artificial Intelligence},
  volume    = {117},
  pages     = {105516},
  year      = {2023},
  publisher = {Elsevier}
}

@article{Manavi2023,
  author    = {Manavi, S. and Becker, T. and Fattahi, E.},
  title     = {Enhanced surrogate modelling of heat conduction problems using physics-informed neural network framework},
  journal   = {International Communications in Heat and Mass Transfer},
  volume    = {142},
  pages     = {106662},
  year      = {2023},
  doi       = {10.1016/j.icheatmasstransfer.2023.106662},
  publisher = {Elsevier}
}

@article{Karniadakis2021,
  author    = {Karniadakis, G. E. and Kevrekidis, I. G. and Lu, L. and Perdikaris, P. and Wang, S. and Yang, L.},
  title     = {Physics-informed machine learning},
  journal   = {Nature Reviews Physics},
  volume    = {3},
  number    = {6},
  pages     = {422--440},
  year      = {2021},
  doi       = {10.1038/s42254-021-00314-5},
  publisher = {Springer Nature}
}

@inproceedings{stipsitz2022approximating,
  author    = {Stipsitz, M. and Sanchis-Alepuz, H.},
  title     = {Approximating the full-field temperature evolution in 3D electronic systems from randomized 'Minecraft' systems},
  booktitle = {ECCOMAS},
  pages     = {31--36},
  year      = {2022},
  publisher = {Scipedia}
}

@article{sabathiel2024,
  author    = {Sabathiel, S. and Sanchis-Alepuz, H. and Wilson, A. S. and others},
  title     = {Neural network-based reconstruction of steady-state temperature systems with unknown material composition},
  journal   = {Scientific Reports},
  volume    = {14},
  pages     = {22265},
  year      = {2024},
  doi       = {10.1038/s41598-024-73380-1}
}

@article{gmsh,
  author    = {Geuzaine, Christophe and Remacle, Jean-Fran{\c{c}}ois},
  title     = {Gmsh: A 3-{D} finite element mesh generator with built-in pre- and post-processing facilities},
  journal   = {International Journal for Numerical Methods in Engineering},
  volume    = {79},
  number    = {11},
  pages     = {1309--1331},
  year      = {2009}
}

@incollection{malinen2013elmer,
  author       = {Malinen, M. and R{\aa}back, P.},
  title        = {Elmer finite element solver for multiphysics and multiscale problems},
  booktitle    = {Multiscale Modelling Methods for Applications in Material Science},
  pages        = {101--113},
  publisher    = {Forschungszentrum Juelich},
  year         = {2013}
}

@inproceedings{helios2022graph,
title = "Towards Real Time Thermal Simulations for Design Optimization using Graph Neural Networks",
author = "H{\`e}lios Sanchis-Alepuz and Monika Stipsitz",
year = "2022",
doi = "10.1109/DMC55175.2022.9906469",
pages = "1--6",
booktitle = "2022 IEEE Design Methodologies Conference (DMC)"
}

@misc{tdk_ntc_datasheet,
  title        = {{TDK} 10K {NTC} Thermistor B57550G1103+005 Datasheet},
  howpublished = {https://docs.rs-online.com/73c1/0900766b813c1e9a.pdf},
  note         = {Data retrieved Feb. 2026},
  organization = {TDK Corporation},
  year         = {2024}
}

@misc{esp32_adc_uncertainty,
  title        = {Cautions in Using ESP32 ADC},
  howpublished = {https://www.makerfabs.cc/article/cautions-in-using-esp32-adc.html},
  note         = {Accessed: Feb. 2026},
  organization = {Makerfabs}
}

@manual{jcgmgum1,
  title        = {{JCGM GUM-1:2023} Guide to the expression of uncertainty in measurement — Part 1: Introduction},
  organization = {Joint Committee for Guides in Metrology (JCGM)},
  year         = {2023},
  url          = {https://www.bipm.org/documents/20126/194484570/JCGM\_GUM-1/74e7aa56-2403-7037-f975-cd6b555b80e6},
  note         = {Accessed: Feb. 2026}
}

@manual{oiml_g101_supp1,
  title        = {{OIML} G 1‑101 (2008) – Evaluation of Measurement Data: Supplement 1 to the “Guide to the Expression of Uncertainty in Measurement” — Propagation of Distributions Using a Monte Carlo Method},
  organization = {International Organization of Legal Metrology (OIML)},
  year         = {2008},
  address      = {Paris, France},
  url          = {https://www.oiml.org/en/files/pdf\_g/g001-101-e08.pdf},
  note         = {Accessed: Feb. 2026}
}

@article{LI2025109393,
title = {PEGNN: A physics embedded graph neural network for out-of-distribution temperature field reconstruction},
journal = {International Journal of Thermal Sciences},
volume = {207},
pages = {109393},
year = {2025},
issn = {1290-0729},
doi={10.1016/j.ijthermalsci.2024.109393},
author = {Qiao Li and Xingchen Li and Xiaoqian Chen and Wen Yao}
}

@article{HU2025125033,
title = {Real-time 3D temperature field reconstruction for aluminum alloy forging die using Swin Transformer integrated deep learning framework},
journal = {Applied Thermal Engineering},
volume = {260},
pages = {125033},
year = {2025},
issn = {1359-4311},
doi = {10.1016/j.applthermaleng.2024.125033},
author = {Zeqi Hu and Yitong Wang and Hongwei Qi and Yongshuo She and Zunpeng Lin and Zhili Hu and Lin Hua and Min Wu and Xunpeng Qin}
}

\end{document}